\documentclass[twocolumn]{aastex63}

\usepackage{natbib}
\usepackage{amssymb}
\usepackage{amsmath}
\usepackage{verbatim}
\usepackage{indentfirst}
\usepackage{xcolor}
\usepackage{graphicx}

\newcommand{\Rnum}[1]{\uppercase\expandafter{\romannumeral #1\relax}}

\newcommand{\OIII}{[\ion{O}{3}]}

\newcommand{\NII}{[\ion{N}{2}]}

\newcommand{\SII}{[\ion{S}{2}]}

\newcommand{\kms}{{km\,s$^{-1}$}}
\newcommand{\Ha}{H$\alpha$}
\newcommand{\ha}{H$\alpha$}

\newcommand{\hb}{H$\beta$}

\newcommand{\civ}{\mbox{C\,\textsc{iv}}}
\newcommand{\mbh}{$M\raisebox{-.3ex}{$\bullet$}$}
\newcommand{\Mbh}{M\raisebox{-.3ex}{$\bullet$}}

\newcommand{\msun}{M$_{\odot}$}

\newcommand{\valerrud}[3]{${#1}^{+#2}_{-#3}$}
\newcommand{\verrud}[3]{{#1}^{+#2}_{-#3}}

\newcommand{\hr}{\mathrm{hr}}

\newcommand{\jav}{\texttt{JAVELIN}}
\newcommand{\zdcf}{{$z$DCF}}

\newcommand{\lris}{{Keck LRIS}}
\newcommand{\gmos}{{Gemini GMOS}}
\newcommand{\siir}{{\SII{} ratio}}

\newcommand{\snu}{\affil{Department of Physics \& Astronomy, Seoul National University, Seoul 08826, Republic of Korea}}
\newcommand{\ucla}{\affil{Department of Physics and Astronomy, University
of California, Los Angeles, CA 90095-1547, USA}}
\newcommand{\ucb}{\affil{Department of Astronomy, University of California, Berkeley, CA 94720-3411, USA}}
\newcommand{\uci}{\affil{Department of Physics and Astronomy, 4129 Frederick Reines Hall, University of California, Irvine, CA 92697, USA}}

\newcommand{\uarizona}{\affil{Department of Astronomy and Steward Observatory, University of Arizona, 933 N. Cherry Ave., Tucson, AZ 85719, USA}}
\newcommand{\umich}{\affil{Department of Astronomy, University of Michigan, Ann Arbor, MI 48109, USA}}
\newcommand{\nasa}{\affil{NASA/GSFC, Code 662, Greenbelt, MD 20771, USA}}

\newcommand{\sdsu}{\affil{Department of Astronomy, San Diego State University, San Diego, CA 92182-1221, USA}}
\newcommand{\knu}{\affil{Major in Astronomy and Atmospheric Sciences, Kyungpook National University, Daegu 41566, Republic of Korea}}
\newcommand{\kasi}{\affil{Korea Astronomy and Space Science Institute, Daejeon 34055, Republic of Korea}}
\newcommand{\nysc}{\affil{National Youth Space Center, Goheung 59567, Republic of Korea}}
\newcommand{\calpoly}{\affil{Physics Department, California Polytechnic State University, San Luis Obispo, CA 93407, USA}}
\newcommand{\cbu}{\affil{Department of Astronomy and Space Science, Chungbuk National University, Cheongju 28644, Republic of Korea}}
\newcommand{\miller}{\affil{Miller Institute for Basic Research in Science, University of California, Berkeley, CA  94720, USA}}
\begin{document}

\title{H$\alpha$ Reverberation Mapping of the Intermediate-Mass Active Galactic Nucleus in NGC 4395}

\author[0000-0003-2010-8521]{Hojin Cho} \snu
\author[0000-0002-8055-5465]{Jong-Hak Woo} \snu
\author[0000-0002-8460-0390]{Tommaso Treu} \ucla
\author[0000-0002-4645-6578]{Peter R. Williams} \ucla

\author{Stephen F. Armen} \sdsu
\author[0000-0002-3026-0562]{Aaron J.\ Barth} \uci
\author[0000-0003-2064-0518]{Vardha N. Bennert} \calpoly
\author[0000-0002-4896-770X]{Wanjin Cho} \snu
\author[0000-0003-3460-0103]{Alexei V. Filippenko} \ucb \miller
\author[0000-0001-5802-6041]{Elena Gallo} \umich
\author{Jaehyuk Geum} \knu
\author[0000-0002-9280-1184]{Diego Gonz\'{a}lez-Buitrago} \uci
\author{Kayhan G\"{u}ltekin} \umich
\author[0000-0002-2397-206X]{Edmund Hodges-Kluck} \nasa
\author{John C. Horst} \sdsu
\author{Seong Hyeon Hwang} \knu
\author{Wonseok Kang} \nysc
\author[0000-0002-3560-0781]{Minjin Kim} \knu
\author{Taewoo Kim} \nysc \cbu
\author[0000-0001-7839-1986]{Douglas C. Leonard} \sdsu
\author[0000-0001-6919-1237]{Matthew A. Malkan} \ucla
\author[0000-0002-0164-8795]{Raymond P. Remigio} \sdsu \uci
\author[0000-0003-4102-380X]{David J. Sand} \uarizona
\author[0000-0001-6363-8069]{Jaejin Shin} \snu \knu
\author{Donghoon Son} \snu
\author[0000-0001-9515-3584]{Hyun-il Sung} \kasi
\author[0000-0002-1912-0024]{Vivian U} \uci
%\author{Stefano Valenti} \lcogt

\correspondingauthor{Jong-Hak Woo}\email{woo@astro.snu.ac.kr}

\begin{abstract}
We present the results of a high-cadence spectroscopic and imaging monitoring campaign of the active galactic nucleus (AGN) of NGC 4395. High signal-to-noise-ratio spectra were obtained at the Gemini-N 8\,m telescope using the GMOS integral field spectrograph (IFS) on 2019 March 7, and at the Keck-I 10\,m telescope using the Low-Resolution Imaging Spectrometer (LRIS) with slitmasks on 2019 March 3 and April 2. Photometric data were obtained with a number of 1\,m-class telescopes during the same nights. The narrow-line region (NLR) is spatially resolved; therefore, its variable contributions to the slit spectra make the standard procedure of relative flux calibration impractical. We demonstrate that spatially-resolved data from the IFS can be effectively used to correct the slit-mask spectral light curves. While we obtained no reliable lag owing to the lack of strong variability pattern in the light curves, we constrain the broad line time lag to be less than 3\,hr, consistent with the photometric lag of $\sim80$\,min reported by Woo et al. (2019). By exploiting the high-quality spectra, we measure the second moment of the broad component of the H$\alpha$ emission line to be $586\pm19$\,{km\,s$^{-1}$}, superseding the lower value reported by Woo et al. (2019). Combining the revised line dispersion and the photometric time lag, we update the black hole mass as $(1.7\pm 0.3)\times10^4$ M$_{\odot}$.
\end{abstract}

\section{Introduction}

Intermediate-mass black holes (IMBHs) with masses between $100$\,\msun{} and $10^{6}$\,\msun{} are perhaps the most elusive astrophysical black holes. Only a small number of them has been reported so far \citep[e.g.,][]{Greene+20}. Detecting IMBHs and carrying out a census of the population is an active subject of research, with potentially far-reaching implications on the formation and growth mechanism of all supermassive black holes \citep[SMBHs; e.g.,][]{Volonteri+03, Volonteri+08, Barai&deGouveiaDalPino19}.

Probing IMBHs via spatially resolved kinematics is extremely difficult since the sphere of influence  ($R_\mathrm{inf}=G \Mbh / \sigma_\ast^2$, where $\Mbh$ and $\sigma_\ast$ denote the black hole mass and the stellar velocity dispersion, respectively) is extremely hard to resolve with current technology. For example, resolution better than $\sim1\,\mathrm{pc}$ is needed for a $10^{5}$\,\msun{} black hole, which is beyond the limit of the {\it Hubble Space Telescope} or adaptive-optics-assisted spectroscopy even for the most nearby galaxies \citep[e.g.,][]{Nguyen+17}.

For active IMBHs, reverberation mapping provides an opportunity to measure the black hole mass by resolving the broad-line region (BLR) through time-monitoring rather than using high-angular-resolution observations. As a first approximation, the black hole mass is given by the time delay ($\tau$) between the continuum and a broad line, in combination with the line width ($\Delta V$), via the formula $\Mbh = f\, c\tau\,(\Delta V)^2 / G$, where $f$ is the virial coefficient \citep{Peterson14}.

NGC 4395 is a nearby dwarf galaxy with a stellar mass of $\sim10^{9}$\,\msun\ \citep{Filippenko&Sargent89, Reines+13}. It contains the least luminous broad-lined active galactic nucleus (AGN) known to date, with a bolometric luminosity lower than $10^{41}\,\mathrm{erg\,s^{-1}}$ \citep{Filippenko&Sargent89, Filippenko+93, Lira+99, Moran+99, Filippenko&Ho03, Cho+20}. While the mass of the central black hole is most likely less than the $10^{6}$ \msun{} limit for IMBHs, its exact mass has been in dispute, with estimates ranging from $9\times10^3$\,\msun{} to $4\times10^{5}$\,\msun{} \citep[e.g.,][]{Filippenko&Ho03, Peterson+05, Edri+12, denBrok+15, Woo+19}. For example, \citet{Peterson+05} measured $(3.6\pm1.1)\times 10^{5}$\,\msun{} based on spectroscopic reverberation mapping of the \civ{} broad emission line, with a time lag of 48--66\,min and a broad-line width of $\sigma \approx 3000\,\mathrm{km\,s^{-1}}$. In contrast, \citet{Woo+19} measured the mass to be $(9.1\pm1.6)\times 10^{3}$\,\msun{} based on the 83\,min time lag obtained from narrow-band photometric reverberation mapping of \ha{} combined with the \ha{} broad-line width of $\sigma = 426$\,\kms{}. Note that the discrepancy between \civ-based and \ha-based black hole masses is mainly due to the dramatic difference of the measured velocity of the two emission lines \citep[see discussion by][]{Woo+19}.

As a follow-up of the previous narrow-band photometric reverberation mapping \citep{Woo+19, Cho+20}, we carried out a high-cadence spectroscopic monitoring campaign of NGC 4395 in 2019, with the goal of resolving this tension and firming up the black hole mass measurement. We observed the AGN over three separate nights, aiming for measuring the short time lag between the AGN continuum and broad Balmer emission lines. We improved traditional slit-based spectrophotometric calibration using the Gemini integral field spectrograph (IFS) data. In this paper, we present the analysis method for flux calibration and the results. While we obtained no conclusive spectroscopic lag of \Ha{} owing to a combination of insufficient variability structure in the light curve and bad-weather losses, we were able to constrain the \ha{} lag to be less than 3 hours. At the same time, the high-quality spectra enabled us to improve the determination of the \ha{} line dispersion and update the black hole mass measurement.

The paper is organized as follows. In Section 2, we describe the observations and data-reduction procedures. Section 3 presentz the data analysis. The revised black hole mass of NGC 4395 is discussed in Section 4, and the results are summarized in Section 5.

\section{Observations and Data Reduction}
Over three nights in 2019, we performed a series of optical spectroscopic observations using GMOS on the Gemini-N 8\,m telescope and the Low-Resolution Imaging Spectrometer \citep[LRIS;]{KeckLRIS, KeckLRIS_Blue} on the Keck-I 10\,m telescope, accompanied by 11 1-2 m class telescopes for photometric observations. Previous studies showed that the BLR size is $\sim 1$--2 light-hours \citep{Peterson+05, Woo+19, Cho+20}, consistent with the extrapolation of the BLR size-luminosity relation \citep{Bentz+13, Cho+20} to low luminosities. In order to measure the time lag, we designed our observations to obtain spectra every 5--10\,min over a full night (7--8\,hr), thus spanning a multiple of the expected lag while sampling it with high cadence. Thanks to multiple telescopes at different longitude, the photometric light curves are more extended in length and have higher cadence than the spectroscopic data.

\begin{figure*}
\centering
\includegraphics[width=0.975\textwidth]{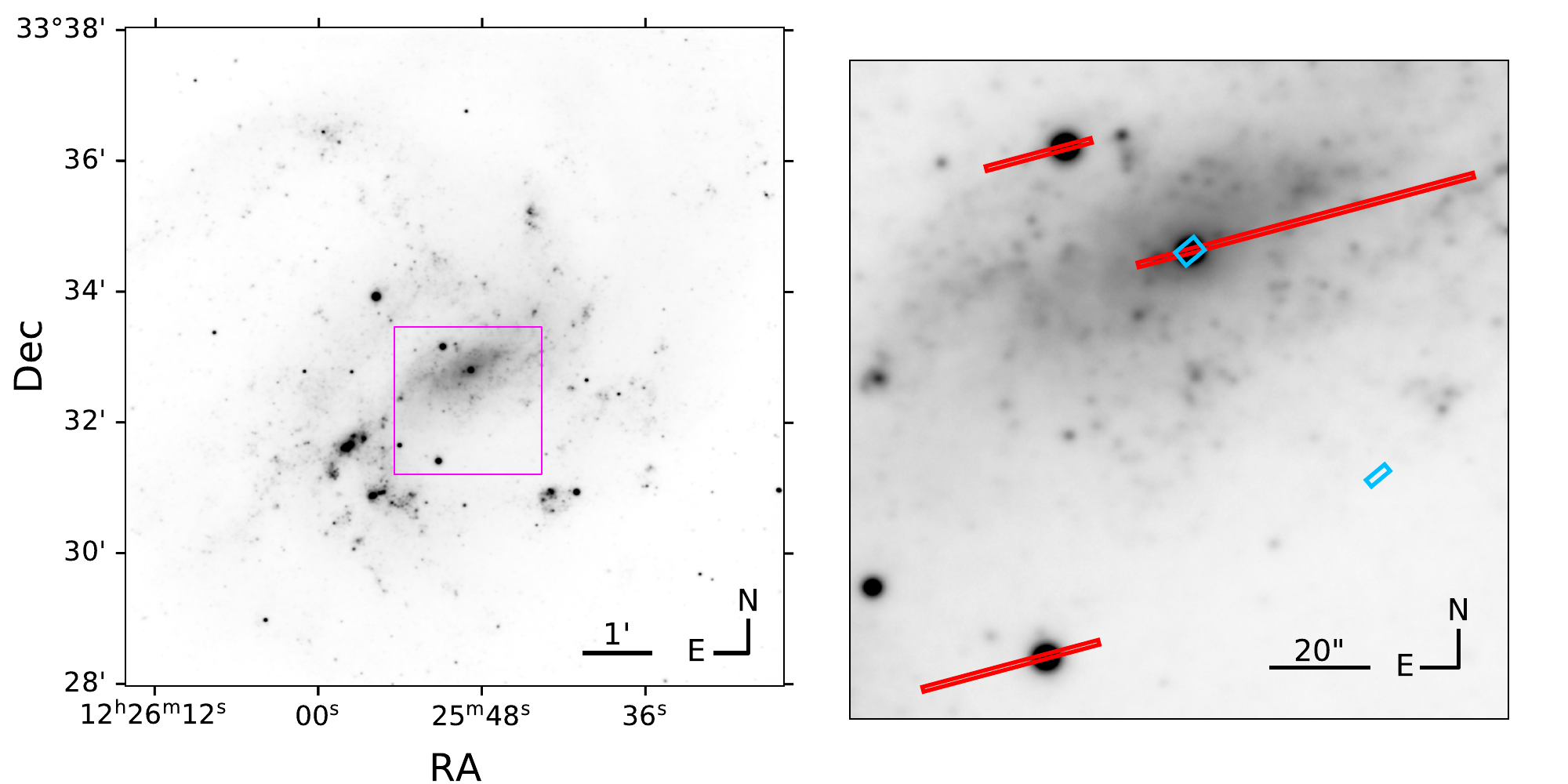}
\caption{\emph{Left}: Example  $V$-band image of NGC 4395, constructed by averaging 140 images with 120\,s exposure each obtained with the Faulkes Telescope North on 3 March 2019. The FoV is $10\arcmin \times 10\arcmin$. The magenta square near the center is magnified in the right panel. \emph{Right}: Magnified image of NGC 4395 with slit masks shown. Red rectangles mark the 1\arcsec{}-wide slits (from west to east, 71\arcsec{} long (the AGN), 22\arcsec{} long (S1), and 37\arcsec{} long (S2) long) used at \lris{} to obtain spectra of the AGN and of the comparison stars, while blue rectangles mark the object field ($3\farcs5 \times 5\arcsec$) and the sky field ($1\farcs75 \times 5\arcsec$) of the \gmos{} IFS.
\label{fig:fov}}
\end{figure*}

\subsection{\gmos}

We observed NGC 4395 on 7 March 2019 (UT dates are used throughout this paper) with \gmos{} (Program ID: GN-2019A-C-2) in integral field unit (IFU-R) mode, which provided a $3\farcs5 \times 5\arcsec$ field of view (FoV); see Fig.~\ref{fig:fov}) and a sufficiently high signal-to-noise ratio (S/N). We used the R831 grating with a central wavelength of 5780\,\AA. In this configuration, the spectral resolving power was $R=7090$, or $\sigma_\mathrm{res}=18\,\mathrm{km\,s^{-1}}$, which was sufficiently high for resolving AGN emission lines. Because of clouds in the early part of the night, only $\sim 5$\,hr of spectroscopic data were obtained. The seeing was $1\arcsec$ at the beginning of the night and improved steadily to  $0\farcs2 $ by the end of the night. The exposure time of individual spectra was 300\,s.

The data were reduced using a modified version of the Py3D pipeline, which is a package for fiber-fed IFU spectrographs originally developed for the reduction of the Calar Alto Large Integral Field Area (CALIFA) survey data \citep{Husemann+13}. The master bias image was constructed by taking the median of 45 associated bias images and then subtracted from all images. Internal flatfield images were used to trace the fibers, and a twilight flatfield was used to correct for differences in the response of each fiber. A wavelength solution was obtained using a Cu-Ar calibration lamp. For each image, a relatively blank and featureless sky region $\sim 1'$ from the nucleus was simultaneously measured and subtracted from a bundle of sky fibers. We used a spectrum of the spectrophotometric standard star HZ~44 to derive the response function and applied it to all epochs obtained throughout the night. Then, each series of fiber spectra was assembled into three-dimensional (3D) datacubes using the drizzle algorithm implemented in Py3D. Atmospheric dispersion \citep{Filippenko82} was corrected by fitting the flux centroid position as a polynomial function of wavelength.

\begin{figure}
\centering
\includegraphics[width=0.45\textwidth]{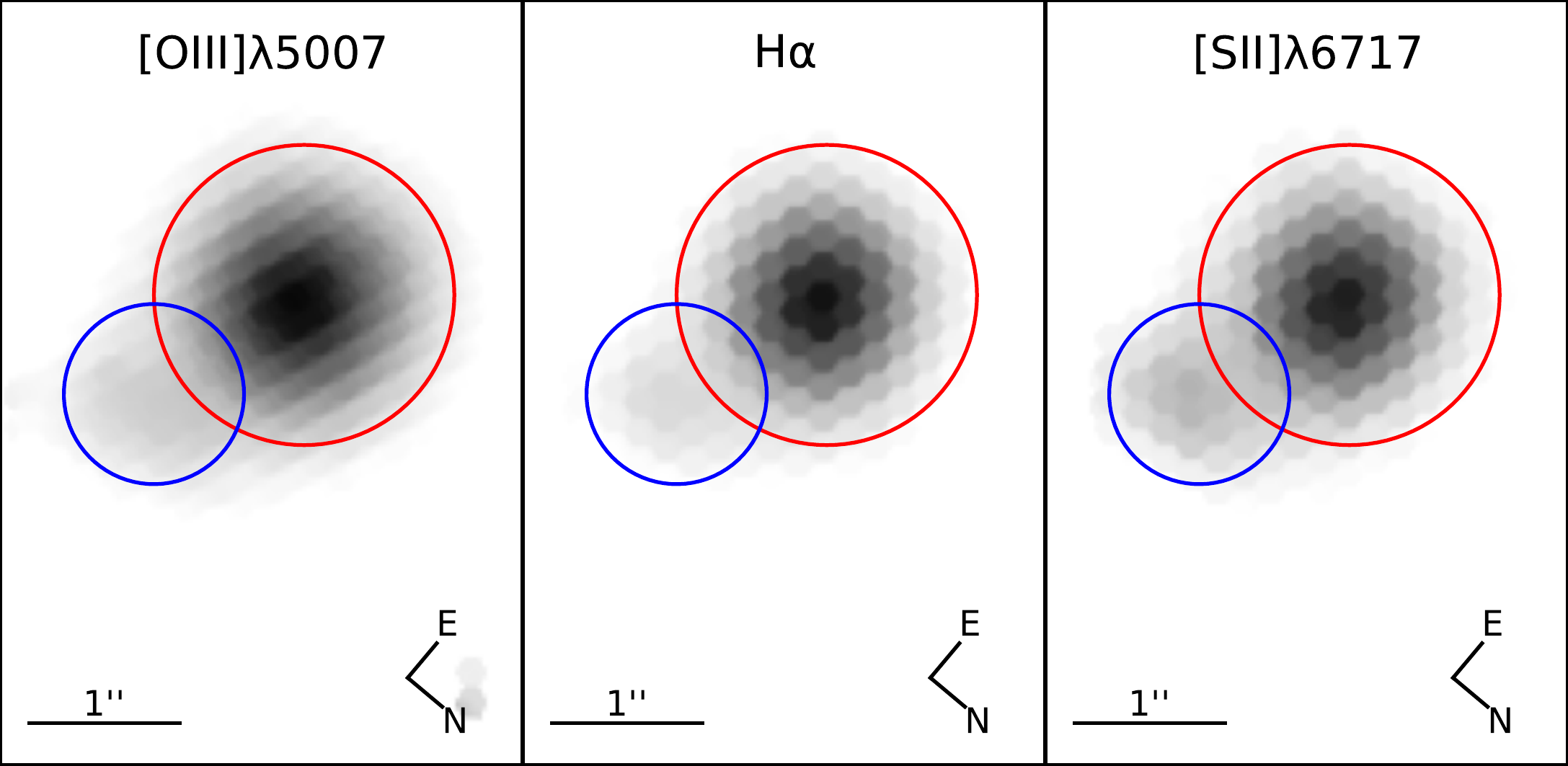}
\caption{Line flux distribution of \OIII{} $\lambda$5007 ({\it left}), \ha{} ({\it middle}), and \SII{} $\lambda 6717$ ({\it right}), which were constructed with GMOS IFU data. Intensity is shown on a logarithmic scale. The size of each panel represents the FoV of the GMOS-IFU (i.e., $3\farcs5 \times 5\arcsec$). Two circles indicate the aperture used for extracting 1-dimensional spectra, one covering the AGN core (red) and the other covering the extended NLR (blue).
\label{fig:refcube}}
\end{figure}

Figure~\ref{fig:refcube} shows GMOS spectral images, which were integrated over the spectral range of each three emission line, after averaging over the data obtained under the best sky conditions. Note that a secondary peak in the line flux distribution was clearly identified at a $\sim$1\arcsec distance from the AGN core, representing an extended narrow-line region (see blue circle in Figure~\ref{fig:refcube}).

For each IFU datacube, a one-dimensional spectrum was extracted within an aperture consisting of two patches, one circle with a radius of 1\arcsec\ centered on the AGN core, and another circle with a radius of $0\farcs6$ at $1\farcs2$ west of the AGN core. The relatively large aperture of an effective area of 4.2 \arcsec$^2$ was used to enclose most of the flux from both the AGN core and the extended NLR, for minimizing the effect of various seeing during the night. An example of the extracted spectrum is presented in Figure~\ref{fig:spec} (bottom panel).

\subsection{\lris}
We observed NGC 4395 on 3 March 2019 and 2 April 2019 using \lris{} with a slit mask in order to include the AGN as well as two comparison stars for secondary flux calibration \citep[similar to the strategy adopted by][]{Williams+20}.  The slits were 1\arcsec{} wide, as shown in Figure~\ref{fig:fov}. On the blue side, we used the 600/4000 grism with $R=1441$. On the red side, we used the 1200/7500 grating, and we measured $R=8522$ or $\sigma_\mathrm{res}=15\,\mathrm{km\,s^{-1}}$ from the night-sky emission lines.

On 3 March 2019, we obtained spectra for 8\,hr under good sky conditions, with a seeing of $\sim1$\arcsec. In contrast, sky conditions were worse with a seeing of $\sim 1\farcs5$ on 2 April 2019, and we were able to acquire only two blocks of exposures of 3\,hr and 1\,hr each, with a pause of 3\,hr between them due to high humidity. For exposure time of individual epochs, we used 369\,s for the red-side CCD and 400\,s for the blue-side CCD in order to match the cadence of the two channels, accounting for the different readout times.

The data were preprocessed using the PypeIt v0.13.0 pipeline \citep{PypeIt}. After bias subtraction, dome flatfield images were used to create pixel flats as well as to trace slits. We derived wavelength solutions from spectra of Ne-Ar lamps on the red side and Hg-Cd-Zn lamps on the blue side. The sky spectra were estimated from the model derived by the PypeIT pipeline to better interpolate the pixels influenced with cosmic rays.

\begin{figure*}[ht!]
\centering
\includegraphics[width=0.9\textwidth]{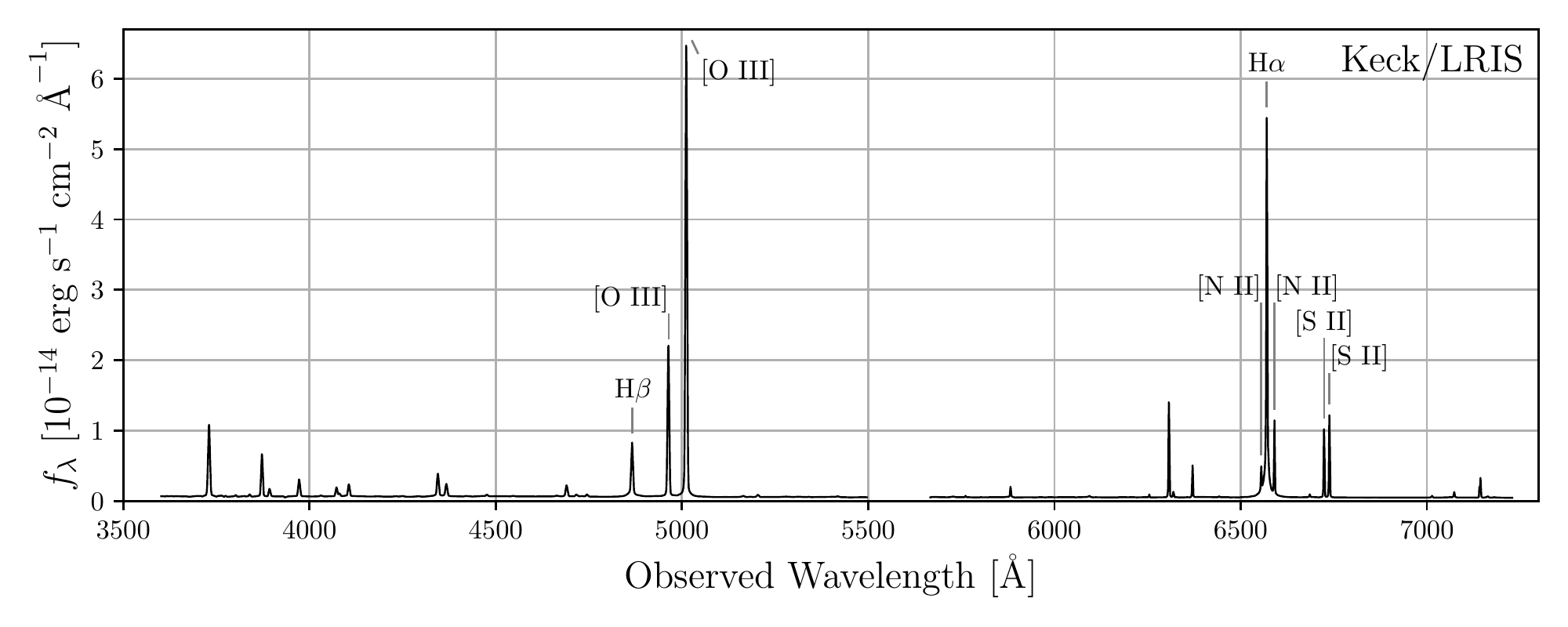}
\includegraphics[width=0.9\textwidth]{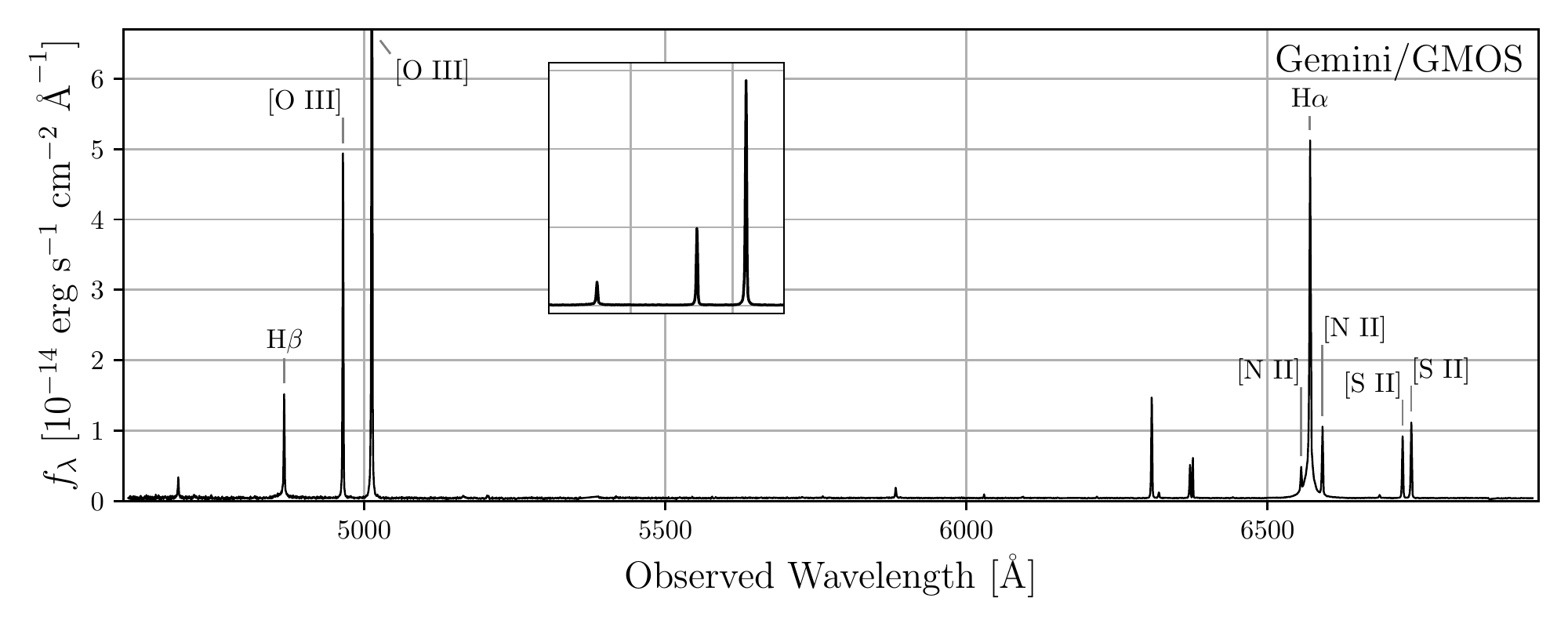}
\caption{Flux-calibrated spectra obtained from {\lris{} on 3 March 2019 ({\it upper panel})} and {\gmos{} on 2 April 2019 ({\it lower panel})}. Note that the \gmos{} spectrum is extracted from a region fully enclosing both the core and the narrow line region as shown in the Figure~\ref{fig:refcube}. As a result, the peak of Balmer lines and the \ha-to-\hb\ flux ratio changes defending on the contribution from the extended NLR, where the \ha-to-\hb\ flux ratio is close to a factor of 3. For clarity, the peaks of \OIII{} and \hb{} emission lines are shown in the inset box.\label{fig:spec}} 
\end{figure*}

Instrumental flexures were manually corrected by comparing the night-sky emission-line centers to the line centers measured by \citet{Cosby+06}. Object spectra, along with the spectra of two bright comparison stars (2MASS J12255090+3333100 (S1) and WISEA J122551.24+333125.7 (S2)), were extracted from the reduced spectral images using an aperture with a width of 1\farcs755. The flux was then calibrated for each exposure by deriving the response function for each spectrum using the comparison stars, as detailed in \S~\ref{sss:keck-fc}. The average spectrum is shown in Figure~\ref{fig:spec} (top panel).

\subsubsection{Flux Calibration}\label{sss:keck-fc}

Since there are no available flux-calibrated spectra of these two comparison stars, we fitted them with spectra taken from a library. First, we estimated the spectral types from their SDSS \emph{ugriz} and 2MASS \emph{JHK} magnitudes using a spectral type fitter \citep[][https://lco.global/{$\sim$}apickles/SpecMatch/]{Pickles&Depagne10}. Then, using spectra in the Indo-US Library of Coud{\'e} Feed Stellar Spectra \citep{Valdes+04} with similar spectral types as templates, we fitted the spectra of S1 and S2 using pPXF \citep{pPXF}. We fitted \ha{} (red side) and \hb{} (blue side) absorption lines and other nearby stellar absorption lines with template spectra using a second-order multiplicative polynomial and without any additive terms. Finally, we determined the best-fitting templates to be HD 165341 (K0\,V) for S1 on the red side and HD 208110 (G0\,IIIs) for S2 on the blue side.

To determine the response functions, we first degraded the template spectra using pPXF to match the resolution of the observed spectra. Then, resolution-matched template spectra were scaled to match the photometric magnitudes obtained from SDSS. Raw spectra of S1 and S2 were divided by the resolution-matched library spectra to obtain quotient spectra. Then, quotient spectra from all epochs were scaled to have the same median values, and the \emph{shape} of the quotients was determined by taking the median of them. Finally, quotient spectra were fitted with the \emph{shape} of quotients multiplied by a low-order polynomial or a power-law function to obtain the response functions. The response functions were applied to the corresponding spectra to produce flux-calibrated spectra of the AGN.

\subsection{Optical Photometric Observations}

In addition to obtaining spectra, we performed photometric observations in the $V$ band to obtain continuum light curves, using Las Cumbres Observatory (LCOGT) telescopes \citep[Teide 0.4\,m, McDonald 0.4\,m \& 1m, and Haleakala 0.4\,m \& Faulkes Telescope North;][]{LCOGT_Brown}, the San Pedro M\'{a}rtir Observatory 0.84\,m telescope, the MDM Observatory 1.3\,m telescope, the Mt. Laguna Observatory 1\,m telescope, the Mt. Lemmon Optical Astronomy Observatory 1\,m telescope, the Bohyunsan Optical Astronomy Observatory 1.8\,m telescope, and the Deokheung Optical Astronomy Observatory 1\,m telescope.  The exposure time for each telescope was set to ensure that the relative photometric error of the AGN is $< 0.02$\,mag, without exceeding 5\,min to preserve the cadence. For each set of observations, we observed Landolt standard-star fields \citep{Landolt92} for absolute flux calibration.

Standard data reduction was performed including bias subtraction and flatfielding using IRAF \citep{Tody86, Tody93, IRAF1999} procedures or \texttt{Ccdproc} package \citep{ccdproc}, and cosmic-ray rejection using the L.A.Cosmic algorithm \citep{vanDokkum+01} as implemented in the \texttt{Astro-SCRAPPY} package \citep{astroscrappy}. After the reduction, data quality was assessed based on visual inspection, and any epoch with quality issues (e.g., failed tracking or performance trouble reported in the observing log) was rejected from further photometric analysis.

\section{Analysis}
\subsection{Decomposition of the broad \ha{} line}\label{ss:decomposition}

\begin{figure}[ht!]
\centering
\includegraphics[width=0.45\textwidth]{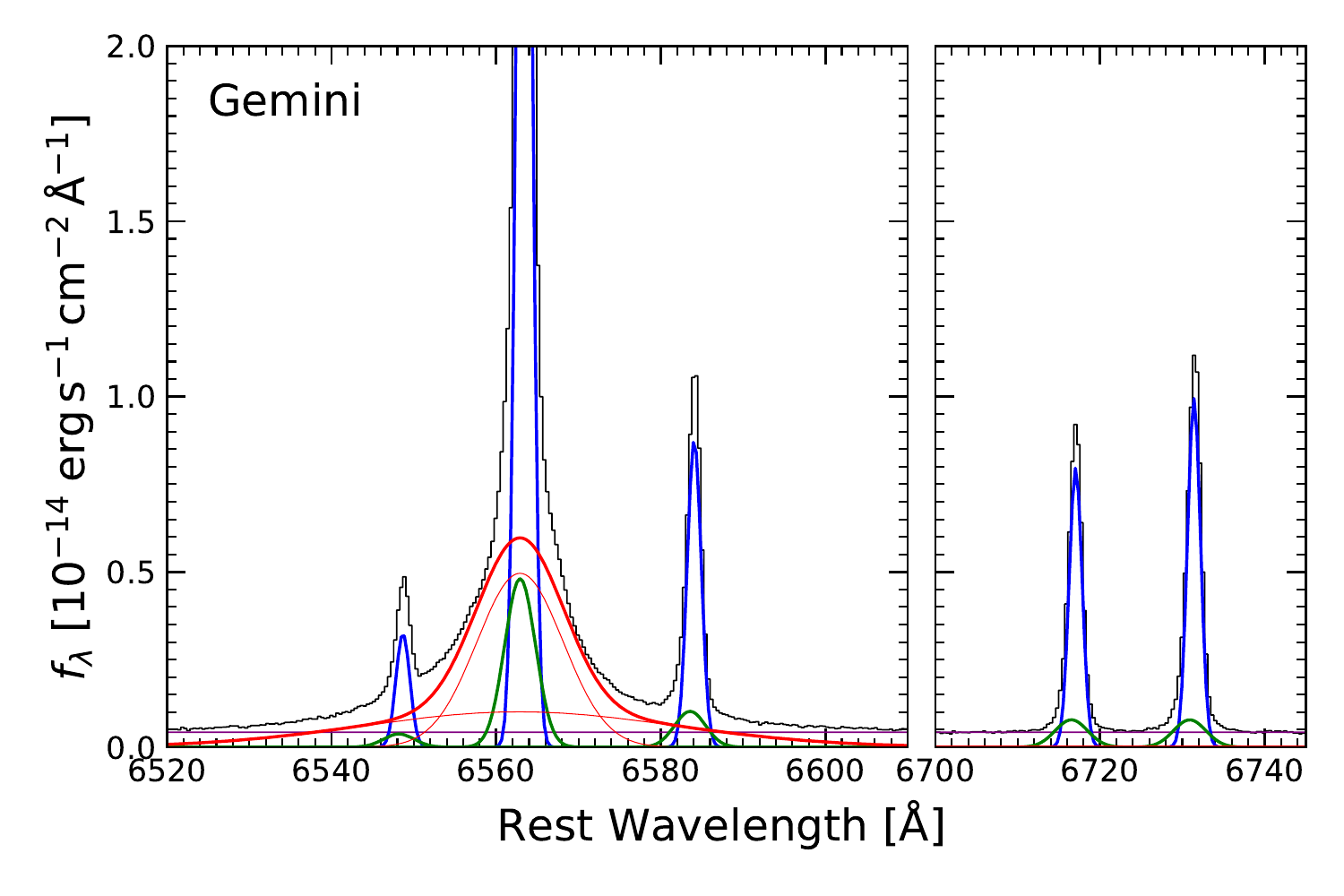}
\includegraphics[width=0.45\textwidth]{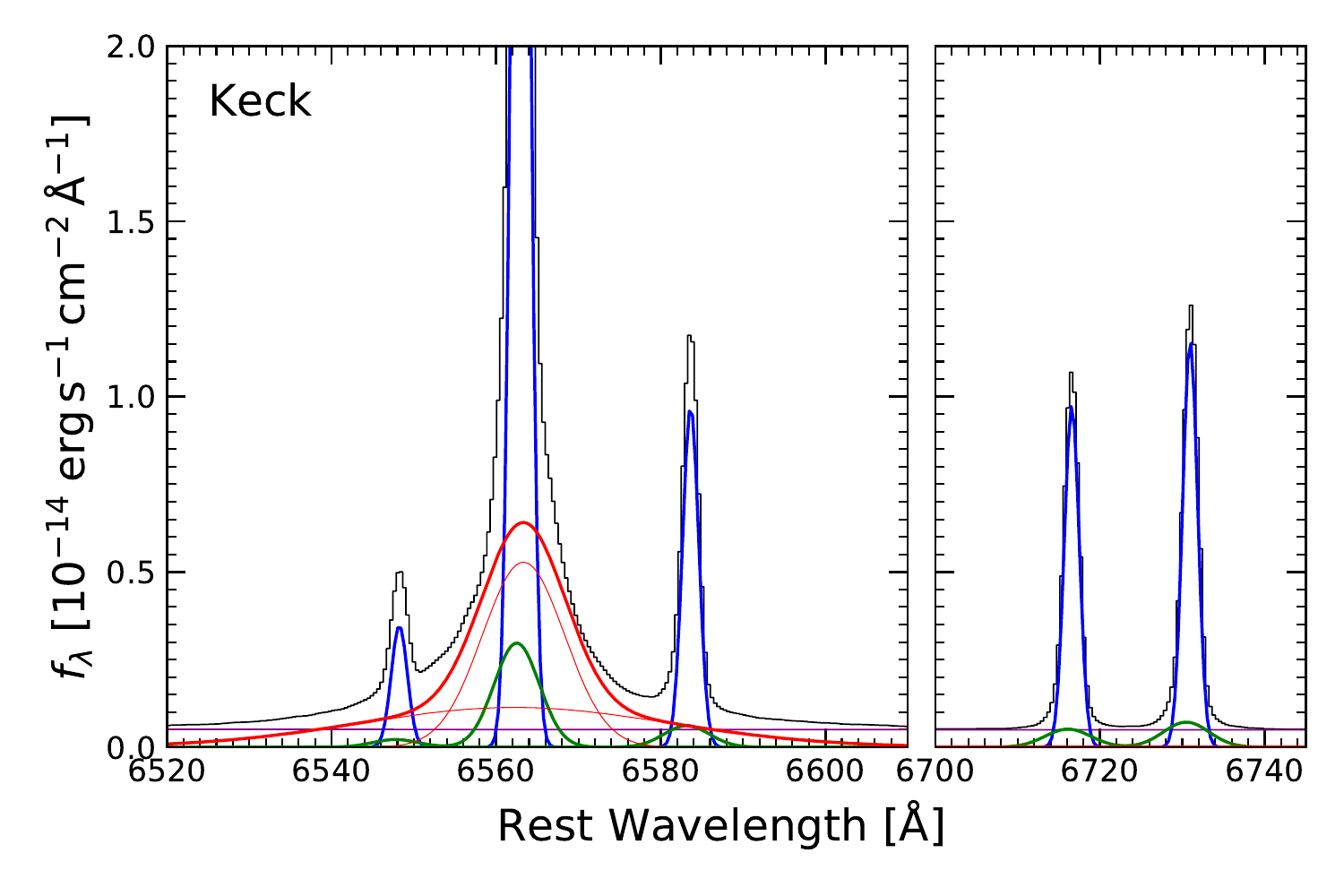}
\caption{Examples of decomposition of \ha{} and \SII{} emission-line regions from \gmos{} (top, on 7 March 2019) and \lris{} (bottom, on 3 March 2019) spectra (black lines). Purple lines represent continuum, blue/green lines denote core/wing components of narrow emission lines, while red thick lines show the broad \ha{}, which is modeled with  two Gaussian components (red thin lines).
		\label{fig:decomposition}}
\end{figure}

To obtain the flux and width of the broad \ha{} emission line, we decomposed the observed spectra as follows. First, we modeled the continuum as a power-law using suitable windows, i.e., 5950--6000\,\AA, 6140--6240\,\AA, 6800--6880\,\AA, and 7220\,\AA{} or redder. Note that a power-law model was chosen because the observed continuum exhibited no significant stellar absorption feature. Second, after continuum subtraction, we modeled the \SII{} $\lambda\lambda$6717, 6731 doublet using two Gaussian components for each line, since both \SII{} lines show a blue-wing component, which represents gas outflows (see right panels of Figure~\ref{fig:decomposition}). Third, we modeled the \ha{} and \NII{} $\lambda\lambda$6548, 6583 lines using two Gaussian components for each narrow line as consistently performed for \SII{}, along with two additional Gaussian components for representing the broad \ha{} line. For \ha{} and \NII{} narrow lines, priors for the line-of-sight (LOS) velocity and the velocity dispersion were taken from the \SII{} line measurements. Examples of emission-line spectra and their decompositions are presented in Figure~\ref{fig:decomposition}. In the case of the broad \ha{} line, we measured the line flux from the best-fit model, and determined the line dispersion by calculating the second moment of the broad-line model, which was constructed with two Gaussian components, after correcting for the instrumental resolution.

For each night, we obtained a mean spectrum by averaging individual spectra taken on that night. For comparison, we report the line dispersion of the broad \ha{} line,
from these mean spectra in Table~\ref{table:linewidth}. We also provide line dispersion of each component in the Gaussian model for both broad \ha{} and narrow lines in Table~\ref{table:linewidth}.
 Overall, the line dispersion of the broad \ha{} ranges from 560 to 600\,km\,s$^{-1}$, of which two Gaussian components have line dispersion of 220--240\,km\,s$^{-1}$ and 800--860\,km\,s$^{-1}$, respectively.

\renewcommand{\arraystretch}{1.25}
\begin{deluxetable*}{lccccccc}[!ht]
    \tablewidth{0.95\textwidth}
    \tablecolumns{8}
    \tablecaption{Line Widths\label{table:linewidth}}
    \tablehead
    {
        \colhead{Instrument}& \colhead{Date (UT)}
        &\multicolumn{3}{c}{{Broad \ha{} Line}}&\multicolumn{3}{c}{{Narrow Lines}}
        \\
        \colhead{}&\colhead{}&\colhead{Comp. 1}&\colhead{Comp. 2}&\colhead{Total}&\colhead{Core}&\colhead{Wing} &\colhead{Total}
        \\
        \colhead{(1)}&\colhead{(2)}&\colhead{(3)}&\colhead{(4)}&\colhead{(5)}&\colhead{(6)}&\colhead{(7)}&\colhead{(8)}
    }
    \startdata
    \lris{}&2019-03-03 &	$226\pm1$ &	$855\pm5$  &	$598\pm3$ &$41.1\pm0.1$ &	$116.5\pm3.8$  &	$59.2\pm1.2$
    \\
    \gmos{}&2019-03-07 &	$229\pm1$ &	$802\pm3$  &	$564\pm2$ &$31.9\pm0.1$ &	$83.0\pm1.6$  &	$48.1\pm0.6$\\
    \lris{}&2019-04-02 &	$240\pm1$ &	$838\pm6$  &	$597\pm4$&$43.0\pm0.1$ &	$111.4\pm1.9$  &	$59.4\pm0.6$
    \enddata
    \tablecomments{Units are in $\mathrm{km\,s^{-1}}$. Uncertainties shown here are 68\% central confidence intervals from the Markov chain Monte Carlo fit posterior.}
\end{deluxetable*}

\subsection{Spectroscopic \ha{} Light Curves}

For the \gmos{} spectra, the response function was measured once throughout the night. Thus, we calibrated the flux of broad \ha{} by assuming constant \SII{} fluxes, following the method by \citet{vanGroningen&Wanders92}. This method is applicable because our apertures were chosen to be large enough to enclose most of the broad and narrow-line emission for the two components. We modeled the \SII{} lines as described in \S~\ref{ss:decomposition}, fixing the difference between the central velocities and the velocity variances of two Gaussian components, as well as the flux ratio between two components, while allowing the center and the width of the core component to be free. We convolved each spectrum with Gaussian kernels to enforce the same spectral resolutions and remove spectral shifts between different exposures, and we scaled the spectra so that the \SII{} lines exhibit exactly the same flux throughout the exposures. We decomposed each spectrum with a power-law continuum, two Gaussians for each narrow line, and two Gaussians for the broad \ha{} component, but we fixed the narrow emission-line kinematics to be the same. Since the \SII{} lines were assumed to have the same flux, the \NII{} and \ha{} narrow lines were also considered to be constant. At the end of these steps, we took the flux of the broad \ha{} components from the decomposition.

In contrast, this method could not be applied to the \lris{} long-slit spectra, because the \SII{} line fluxes showed significant variability in their ratio, even after flux calibration using the comparison stars. We attribute this variability to contamination from the secondary peak of narrow-line emission, which contributes a different amount to each exposure based on the seeing conditions and the mask alignment. Owing to the lack of stable narrow lines to be used as spectral calibrators, we could not apply the \citet{vanGroningen&Wanders92} method. Thus, after we modeled and subtracted the continua as described in \S~\ref{ss:decomposition}, the flux within the wavelength range 6480--6640\,\AA{} in the rest frame is summed to estimate the total flux of broad \ha{} and narrow \ha{} and \NII. Finally, we used the method described in \S~\ref{sss:nl_excess} to separate the narrow-line flux from the broad-line flux.

\subsubsection{Narrow-Line Variability of \lris{} Spectra}\label{sss:nl_excess}

\begin{figure*}[ht!]
\centering
\includegraphics[width=0.54\textwidth]{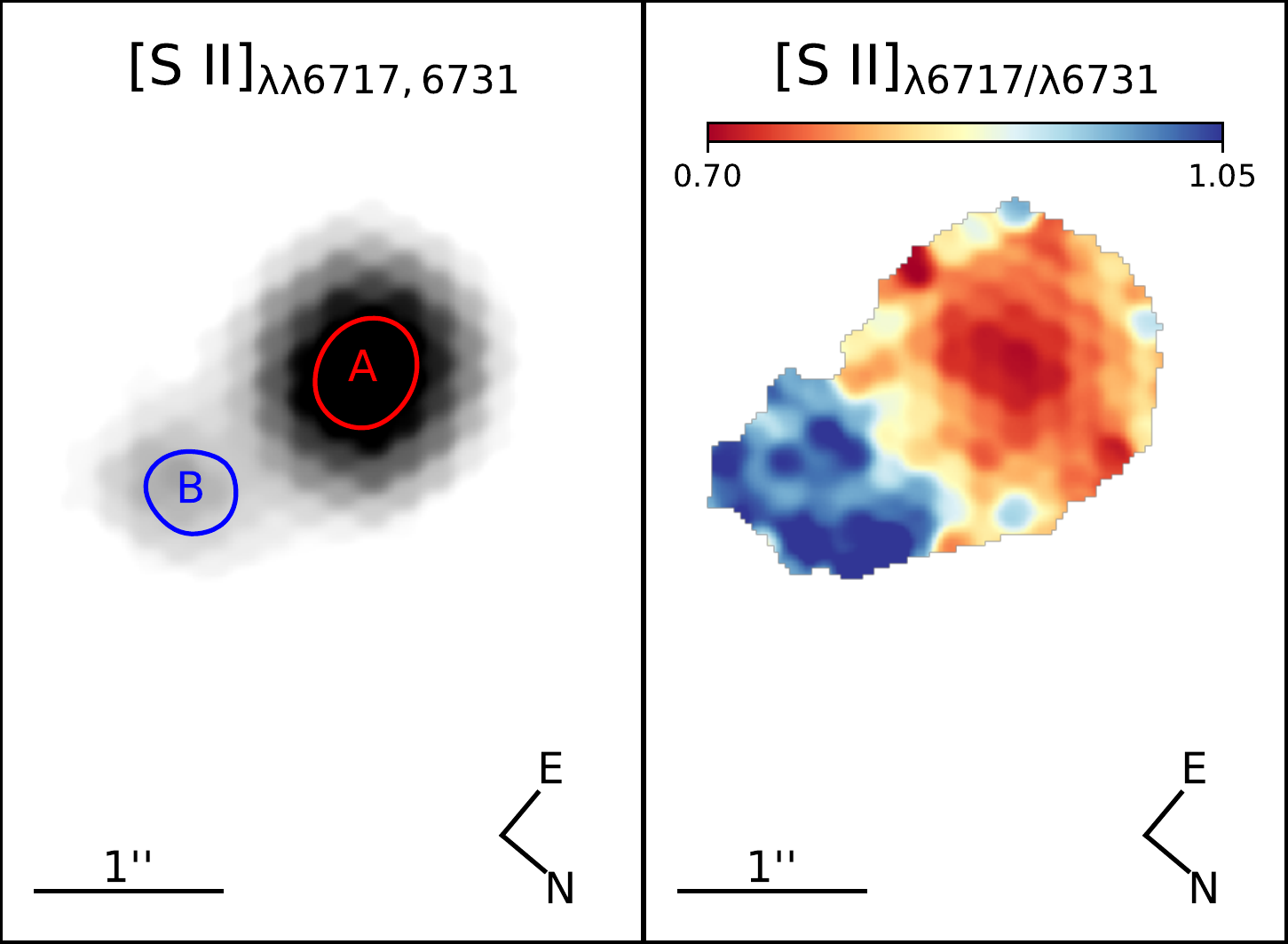}
\includegraphics[width=0.405\textwidth]{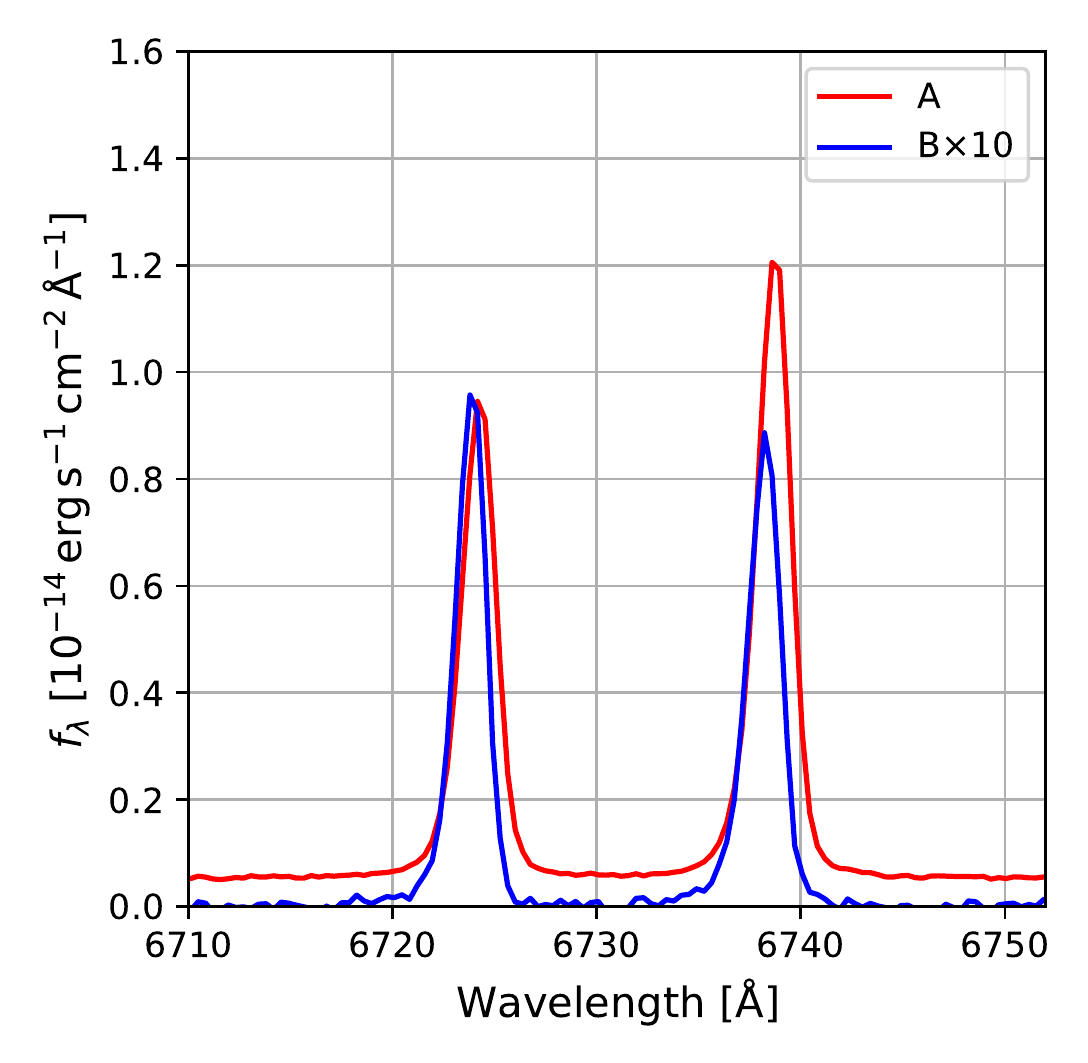}
\caption{Differences in the \SII{} line ratio between the core (A) and the NLR secondary peak (B).
{\it Left:} Map of the \SII{} doublet flux and its flux ratio.
{\it Right:} The line profile of the \SII{} doublet from the core (red) and the secondary peak (blue), with appropriate scaling.  The two regions show different \SII{} $\lambda 6717/\lambda 6731$ ratios as well as different line-of-sight velocities.
\label{fig:sii_blob}}
\end{figure*}

Figure~\ref{fig:sii_blob} shows the spatial distribution of \SII{} $\lambda 6717/\lambda 6731$ ratios (hereafter \siir{}) from the GMOS IFU data. It is clear that the central part of the AGN has a significantly lower \siir{} compared to the secondary peak of the NLR to the west. Thus, as observed, we expect changes in narrow-line flux due to contamination from the extended NLR to be accompanied by changes in \siir{}. We use the following approach to disentangle the broad \ha{} flux from the narrow lines.

First, we assume that the observed spectrum is a mixture of two distinct spectra from the core and the secondary peak.  If the flux calibration based on comparison stars were perfect, we could describe the expected total \SII{} flux ($F$) as a function of the \siir{} ($\beta$),
\begin{align}
F(\beta) = F_{0}\;\frac{1+\beta}{1+\beta_\mathrm{A}} \cdot \frac{\beta_\mathrm{B} - \beta_\mathrm{A}}{\beta_\mathrm{B}-\beta},
\end{align}
where $\beta_\mathrm{A} = 0.75$ and $\beta_\mathrm{B} = 1.02$ are obtained from the spectra extracted from the regions specified in Figure~\ref{fig:sii_blob}. The flux of \SII{} lines at the core, $F_{0}$, can be determined by fitting the data.

\begin{figure}[ht!]
\centering
\includegraphics[width=0.45\textwidth]{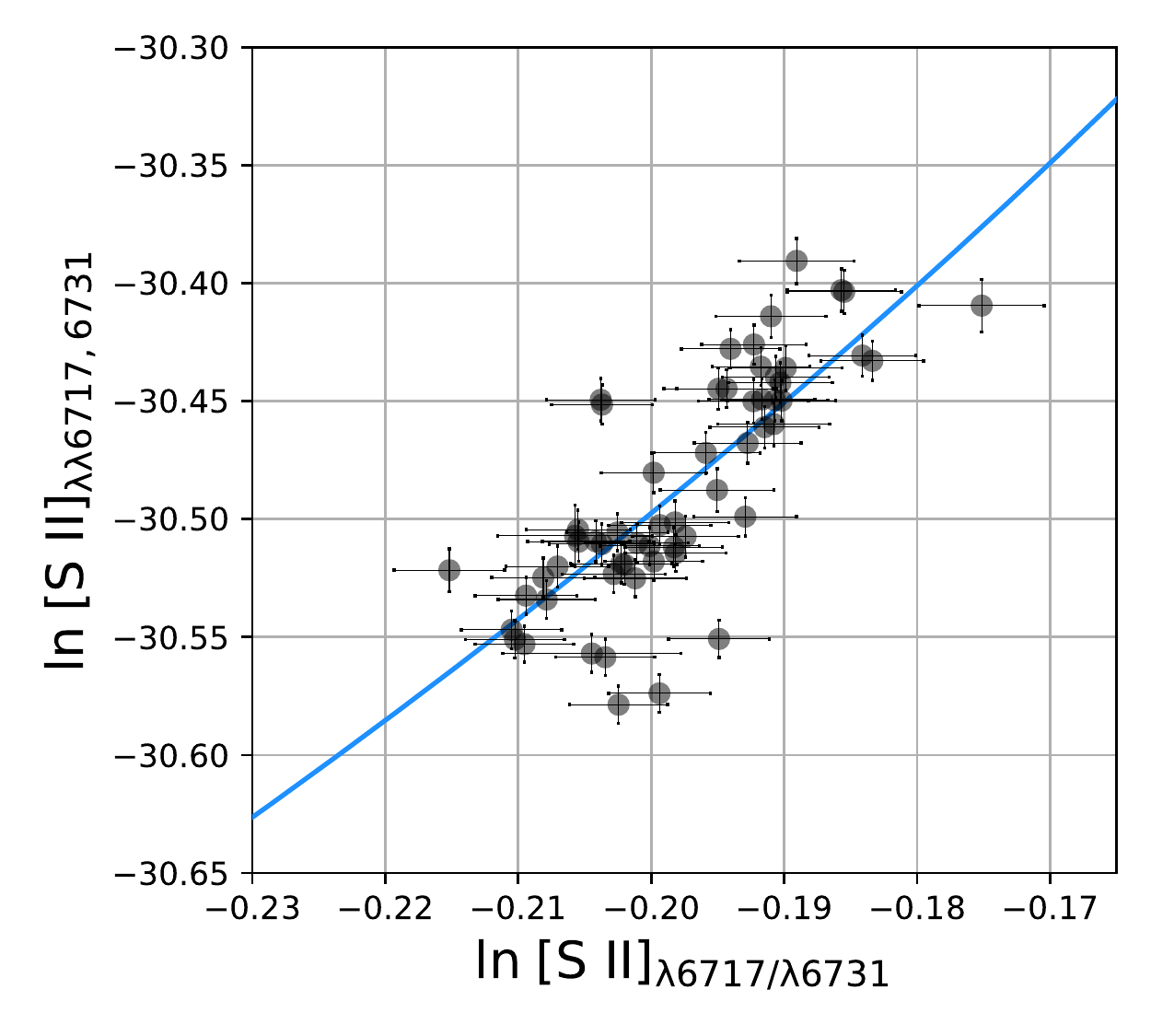}
\caption{
Observed \SII{} line flux against the \SII{} line ratio from \lris{}. The abscissa represents the line ratio of the \SII{} $\lambda 6717/\lambda 6731$ doublet, and the ordinate represents the total flux of \SII\ $\lambda\lambda6717$, 6731 in ${\rm erg\,s^{-1}\,cm^{-2}}$; both are on logarithmic scales. Each point represents a measurement from each spectrum obtained on 3 March 2019. The blue curve shows our model for the relationship between them, assuming the narrow-line flux variation is completely explained by the flux of the nearby NLR secondary peak seeping into the slit. The shape of the curve is determined from the GMOS IFU analysis, and the flux scale ($y$-intercept in the plot) is obtained by minimum $\chi^2$ fitting, with $\chi_{\nu}^2 = 1.12$ for the best-fit model.
\label{fig:sii_ratio_fit}
}
\end{figure}

The result is presented in Figure~\ref{fig:sii_ratio_fit}. The reduced $\chi_{\nu}^2=1.12$ suggests that the model explains the data well. This supports our hypothesis that the flux variation in narrow lines is indeed due to the flux of the NLR secondary peak seeping into the extraction aperture differently for each epoch. We thus conclude that the flux and the line profile of narrow \ha{} and \NII{} changed during the observations owing to the varying contribution from the secondary peak, rendering the \ha{} decomposition inaccurate.

\begin{figure}[ht!]
\centering
\includegraphics[width=0.45\textwidth]{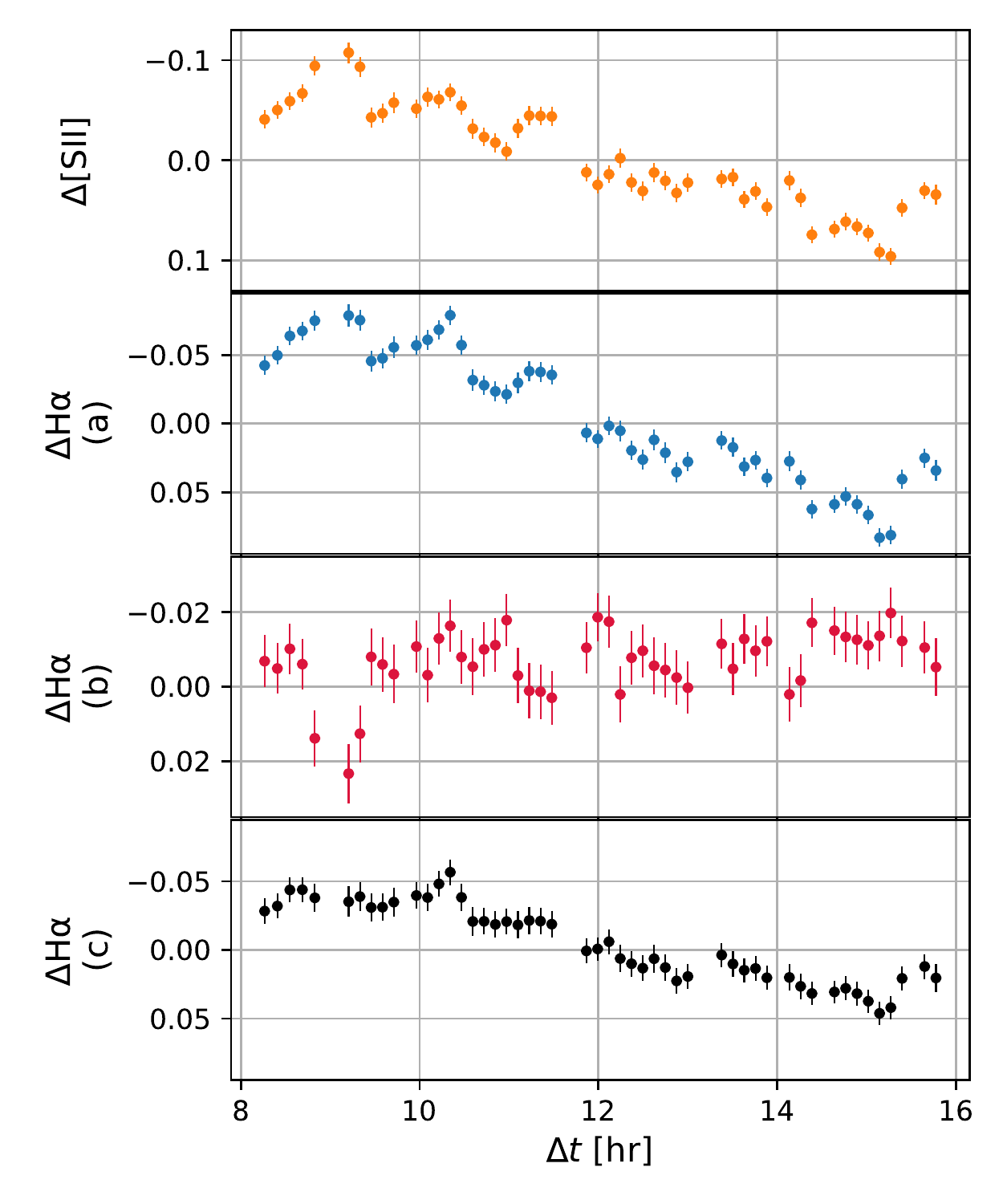}
\caption{
Correction of \ha{} flux using the \SII{} flux. The top panel shows the flux of \SII{} lines (yellow), and the bottom three panels show the measured total flux of \ha{} and \NII{} lines, both narrow and broad, in magnitude difference units. (a) The raw total flux of \ha{} and \NII{} (blue). (b) Total flux of \ha{} and \NII{} corrected by the \citet{vanGroningen&Wanders92} scheme --- i.e., the total raw flux of \ha{} and \NII{} divided by the \SII{} flux (red). (c) Total flux of \ha{} and \NII{} after subtracting excess narrow-line flux, which is calculated from the \SII{} line flux excess (black). This is based on the assumption that the flux calibration is reliable, and consequently that the observed narrow-line fluxes are not constant. We argue (c) to be correct, as demonstrated in \S~\ref{sss:nl_excess}.
\label{fig:sii_correction}}
\end{figure}

Figure~\ref{fig:sii_correction} demonstrates our flux-correction method to compensate for the excess narrow-line fluxes and isolate the broad \ha{} light curve. First, we estimated the \SII $\lambda\lambda$6717, 6731 excess by subtracting the model \SII{} flux \mbox{$F(\beta=0.75)=2.28\times10^{-14}\,\mathrm{erg\,s^{-1}\,cm^{-2}}$} from the measured \SII{} flux. Then, we measured the total flux of the \NII{} and \ha{} narrow lines in the NLR secondary peak to be $2.33\pm 0.01$ times that of \SII{} fluxes and multiplied it by the \SII{} excess to calculate the \ha{} and \NII{} excess flux for each spectrum. Finally, we calculated the total flux of \ha{} and \NII{} (including both narrow and broad lines) for each spectrum and subtracted the excess \NII{} and \ha{} flux, the result of which should only contain the flux of broad \ha{} and invariant narrow lines. Note that the shape of the corrected light curve would be drastically different if we followed the recipe of \citet{vanGroningen&Wanders92}.

\subsection{Differential Photometry}

We follow the method demonstrated by \citet{Cho+20} to obtain the continuum flux light curves. We first constructed point-spread functions (PSFs) for each image based on isolated, bright, and unsaturated stars. Second, we matched the resolution of the images taken during each night by convolving the sharpest images with a suitable kernel.

Third, we performed aperture photometry of the AGN and nearby comparison stars on the PSF-matched images using the \texttt{photutils} package \citep{photutils07}. We estimated and subtracted the global sky background using the \texttt{SExtractorBackground} estimator implemented in the  \texttt{photutils} package. For each source, we estimated the residual sky background from annuli with innerand outer radii of (respectively) 2--3 and 3--5 times the seeing full width at half-maximum intensity (FWHM) to subtract from the signal. We determined the aperture size so as to maximize the S/N of the integrated AGN flux (typically $\sim 2$ times the seeing FWHM).  The aperture size of comparison stars was fixed to be the same as that of the AGN.

To obtain the relative AGN light curves, we calculated offsets of the instrumental magnitudes of comparison stars ($\delta\mathrm{V}_{\mathrm{C}i}$), and adopted the mean offset for each image as the relative normalization value ($\delta\mathrm{V}=\left\langle\delta\mathrm{V}_{\mathrm{C}i}\right\rangle_i$). The variances of the offsets were added as systematic uncertainties to estimate the total uncertainties of the AGN light curve.
Finally, the light curves of different telescopes for the same night were
intercalibrated by adding a linear shift in the magnitudes so that they share the same average magnitude within the overlapping portion of the light curve. We only used sets of the light curves with average magnitude uncertainties below 0.02\,mag for the time-lag measurements.

\subsection{Time-Lag Measurements}

\begin{figure}[!t]
\centering
\includegraphics[width=0.45\textwidth]{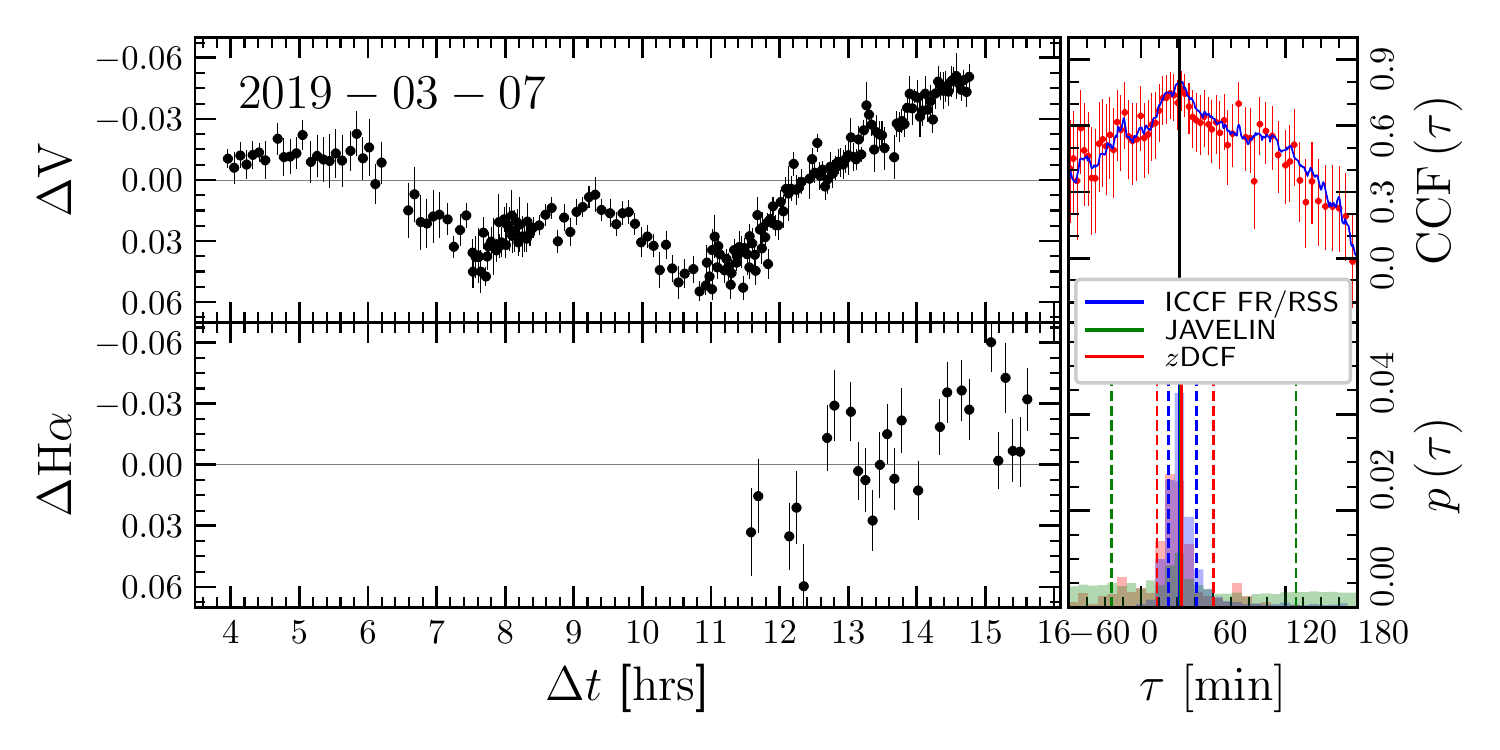}
\caption{Light curves and time-lag measurement using Gemini/GMOS on 7 March 2019. Left panels show the light curves from $V$ (upper left) and \ha{} (lower left). Right panels show the CCF (upper right) and the time-lag measurements (lower right). ICCF, zDCF, and \jav{} are indicated as blue,
red, and green colors, respectively. Vertical solid lines represent the median of each measurement, while dashed lines mark 16\%-ile and 84\%-ile as uncertainties. \label{fig:gmosccf}}

%\end{figure}
%\begin{figure}[!t]
%\centering
\includegraphics[width=0.45\textwidth]{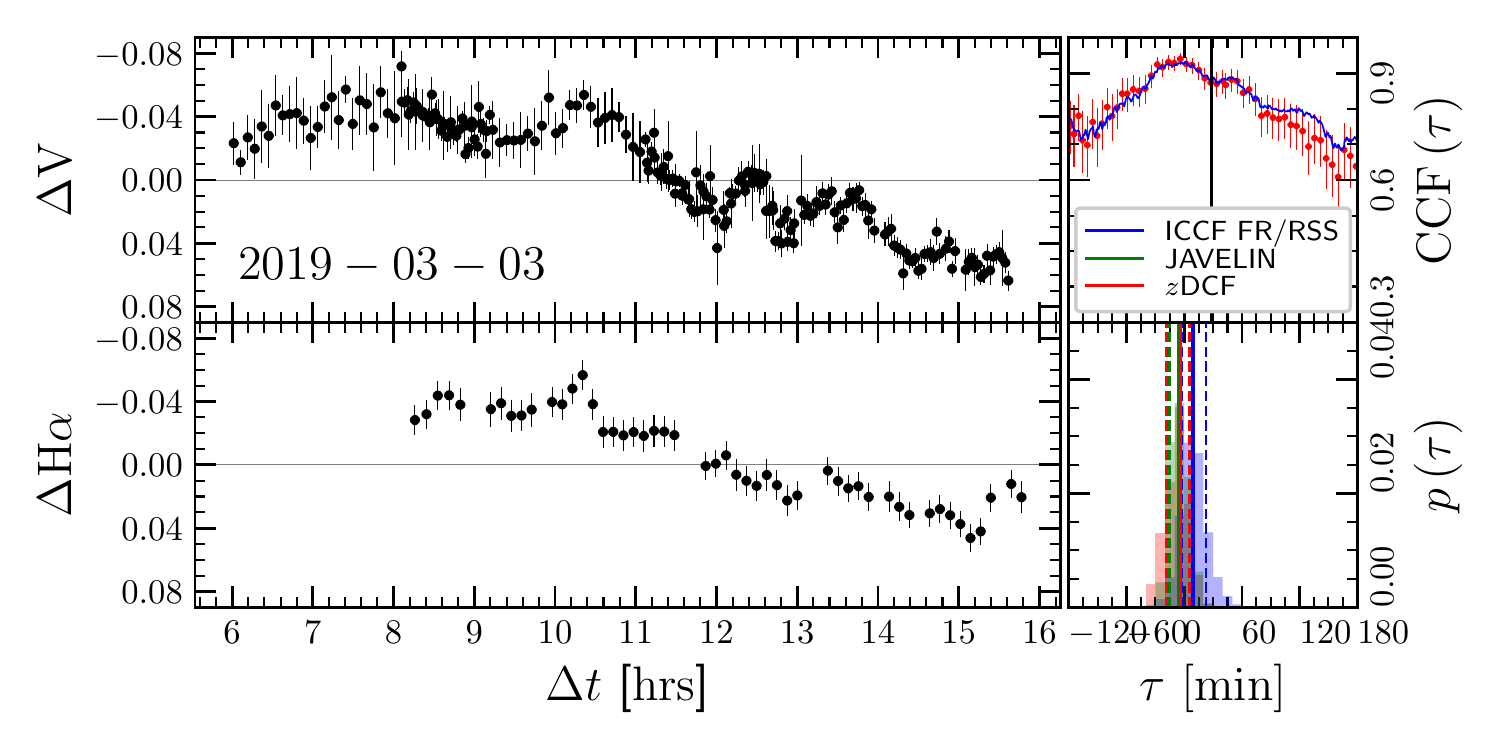}
\caption{Light curves and time-lag measurement using Keck/LRIS on 3 March 2019. Left panels show the light curves from $V$ (upper left) and \ha{} (lower left). Right panels show the CCF (upper right) and the time-lag measurements (lower right). ICCF, zDCF, and \jav{} are indicated as blue, red, and green colors, respectively. Vertical solid lines represent the median of each measurement, while dashed lines mark 16\%-ile and 84\%-ile as uncertainties.\label{fig:keck1ccf}}

%\end{figure}
%\begin{figure}[!b]
%\centering
\includegraphics[width=0.45\textwidth]{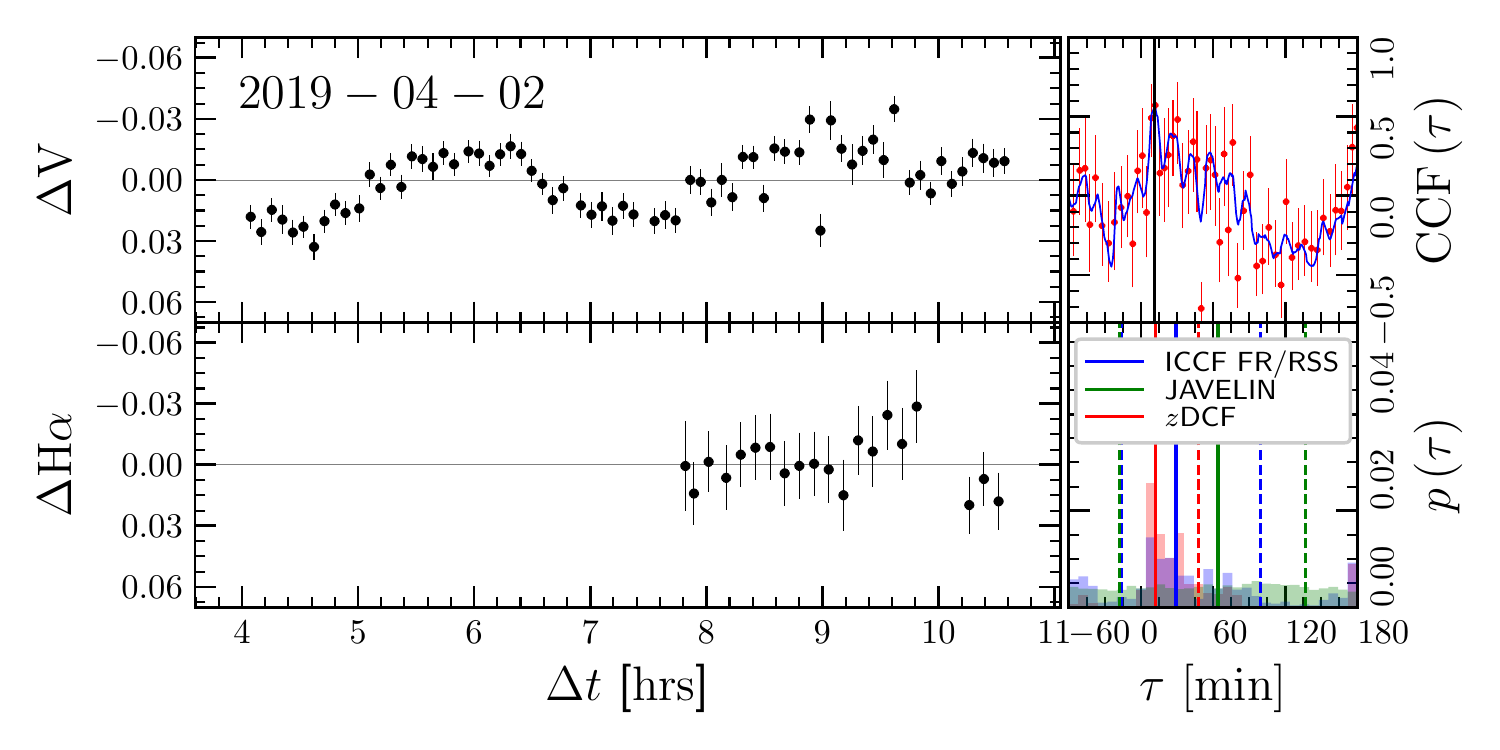}
\caption{Light curves and time-lag measurement using Keck/LRIS on 2 April
2019. Left panels show the light curves from $V$ (upper left) and \ha{} (lower left). Right panels show the CCF (upper right) and the time-lag measurements (lower right). ICCF, zDCF, and \jav{} are indicated as blue, red, and green colors, respectively. Vertical solid lines represent the median of each measurement, while dashed lines mark 16\%-ile and 84\%-ile as
uncertainties.\label{fig:keck2ccf}}
\end{figure}

\begin{figure*}[!t]
\centering
\includegraphics[width=0.95\textwidth]{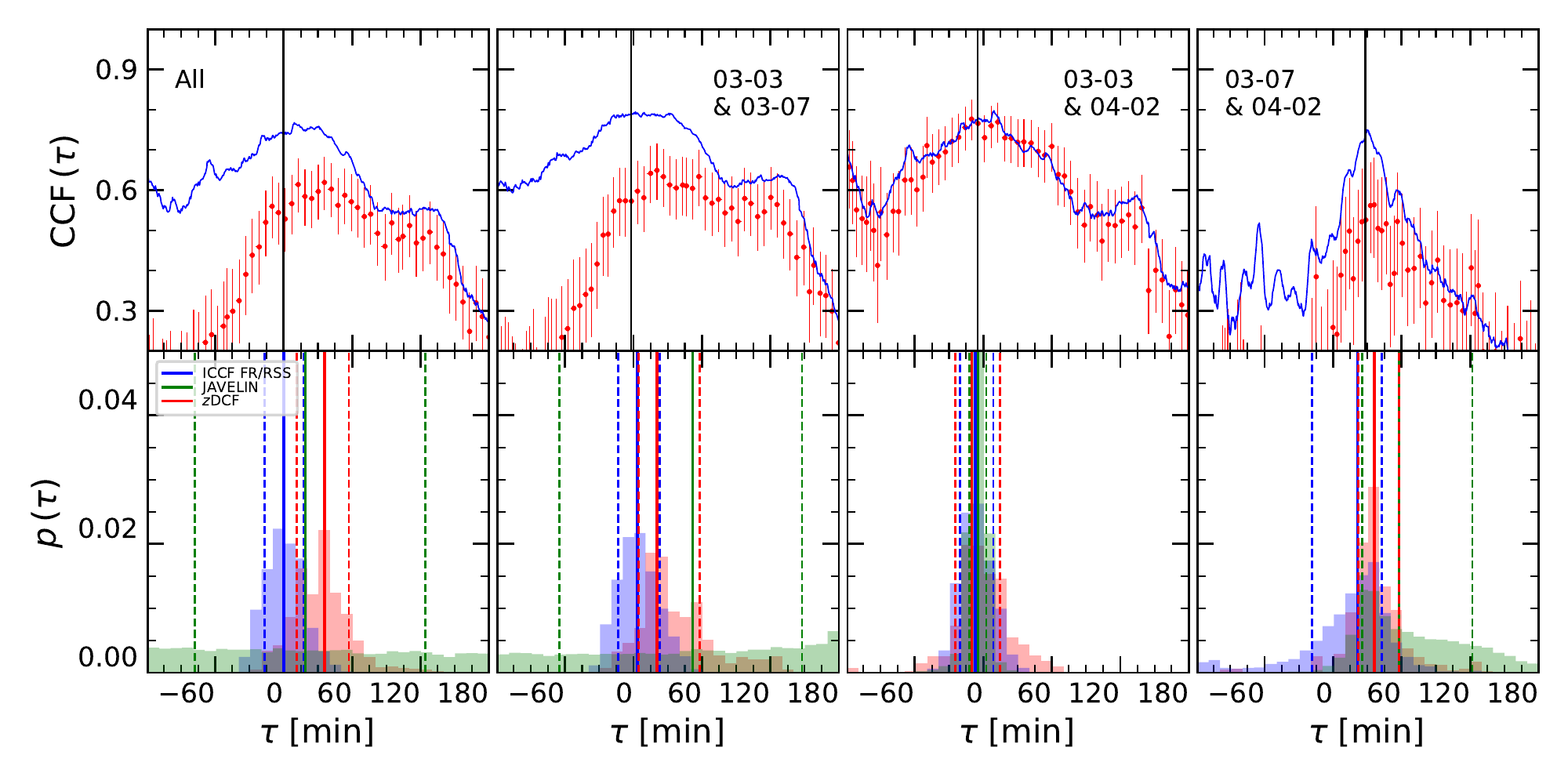}
\caption{Time-lag measurements using combinations of observed light curves. Top panels show the CCF, while bottom panels show the time0lag measurements. ICCF, zDCF, and \jav{} are indicated as blue, red, and green colors, respectively. Four sets of panels show the result when combining all light curves, all but 2019-04-02, all but 2019-03-07, and all but 2019-03-03 (from left to right). Vertical solid lines represent the median of each measurement, while dashed lines mark 16\%-ile and 84\%-ile as uncertainties.\label{fig:combccf}}
\end{figure*}

The time lag between the $V$ and \Ha{} light curves was computed following the method described by \citet{Cho+20}.
For each pair of light curves, we computed the interpolated cross-correlation function (ICCF; \citealt{White&Peterson94}), with the flux randomization/random-subset selection method (FR/RSS; \citealt{Peterson+98}; see also \citealt{Peterson+04}) to estimate its uncertainty. The ICCF $r(\tau)$ was computed over $-2\,\hr<\tau<3\,\hr$ for the 3 March 2019 observation with a long light curve, and over $-1\,\hr<\tau<3\,\hr$ for other two nights. ICCF centroids were calculated over an interval ($I$) containing the ICCF peak ($r(\tau_{\mathrm{peak}}\in I)= r_\mathrm{max}$) with $r(\tau\in I)>0.8\, r_\mathrm{max}$. We simulated 2000 realizations with the FR/RSS, and the median and the lower/upper bounds of the 68\% central confidence interval of the centroid distribution were taken as the time lag and its lower/upper uncertainty. We also used the $z$-transformed discrete correlation function (\zdcf;  \citealt{Alexander97}) and the \jav\ method \citep{Zu+11} to check the consistency with the ICCF results. The measurements are summarized in Table~\ref{table:timelag}, and CCFs and time-lag measurements are presented in Figures~\ref{fig:gmosccf}, \ref{fig:keck1ccf}, \ref{fig:keck2ccf}, and \ref{fig:combccf}.

\renewcommand{\arraystretch}{1.25}
\begin{deluxetable}{crrr}[!t]
	\tablewidth{0.4\textwidth}
	\tablecolumns{4}
	\tablecaption{\ha{} Time-Lag Measurements\label{table:timelag}}
	\tablehead
	{
		\colhead{Date (UT)}&\colhead{ICCF}&\colhead{\zdcf}&\colhead{\jav}
		\\
		\colhead{(1)}    &\colhead{(2)}       &\colhead{(3)}       &\colhead{(4)}
	}
	\startdata
	2019-03-03 &	\valerrud{9}{13}{12} min. & \valerrud{-4}{10}{14} min. & \valerrud{-6}{11}{9} min.\\
	2019-03-07 &	\valerrud{32}{14}{9} min. & \valerrud{34}{26}{20} min. & \valerrud{33}{96}{57} min.\\
	2019-04-02 &	\valerrud{29}{70}{45} min.& \valerrud{12}{36}{76} min. & \valerrud{64}{73}{82} min.\\
	\hline
	03-03 \& 03-07 &	\valerrud{3}{20}{17} min.& \valerrud{21}{37}{16} min. &
\valerrud{52}{96}{116} min.\\
	03-03 \& 04-02 &	\valerrud{-7}{16}{13} min.& \valerrud{-10}{25}{14} min.
& \valerrud{-5}{7}{8} min.\\
	03-07 \& 04-02 &	\valerrud{21}{21}{40} min.& \valerrud{36}{22}{14} min. & \valerrud{58}{64}{32} min.\\
	All &	\valerrud{0}{18}{17} min.& \valerrud{36}{21}{24} min. & \valerrud{19}{105}{97} min.
	\enddata
    \tablecomments{Uncertainties shown here are 68\% central confidence interval.}
\end{deluxetable}

We did not conclusively measure a time lag in the 3 March 2019 observation even though it had the longest time baseline with a significant variability amplitude. This is because the AGN variability pattern during that night was mostly a monotonic decrease in flux, without a strong pattern or inflection points, which could help reducing the uncertainty of the time lag in the cross-correlation analysis.
In the case with the data taken on  7 March 2019 and 2 April 2019, we only tentatively measured the time lag between $V$ and \ha{} to be $\sim 30$\,min with larger uncertainty. Note that due to bad weather the time base line is short and the number of spectroscopic epochs is relatively small in the \ha{} light curve, which aggravated the cross-correlation analysis. When we combined the light curves obtained during different nights, the improvement of the lag measurement was not significant. Therefore, we conclude that these lag measurements are not reliable, and additional monitoring is needed to verify it.

On the other hand, we note that both the ICCF and \zdcf{} reduce to negligible numbers in all of the combinations above when the time lag is close to 3\,hr. This, combined with the shape of the light curves, suggests that the time lag cannot be longer than 3\,hr.

\section{Discussion}

\subsection{Broad-Line Width and AGN Mass}

We determine the mean width of broad \ha{} as $\sigma_\mathrm{BH\alpha} = 586\pm19\,\mathrm{km\,s^{-1}}$, by taking the average and standard deviation of individual measurements from each of 3 nights (see Table~\ref{table:linewidth}). This is 36\% broader than the measurement reported by \citet{Woo+19}, who used a single Gaussian for fitting the broad \ha{} (i.e., $\sigma=426\pm1\,\mathrm{km\,s^{-1}}$). Note that the difference in the decomposition model is responsible for the difference of the line dispersion measurement of broad \ha{}. Specifically, in \citet{Woo+19} the continuum near \ha{} was estimated as a linear function over a short range of wavelengths, whereas we modeled the continuum as a power law using a much wider range of wavelengths. A linear fit can overestimate the power-law-like continuum near \ha{}, which leads to a reduced flux in the wing of the broad \ha{} line. A reanalysis of the spectrum presented in that paper using the methods described here yields $\sigma_\mathrm{BH\alpha} = 592$~\kms{}. This is in
excellent agreement with the values measured with the newly obtained spectra, and the Bayesian information criterion as well as the $\chi_{\nu}^2$ of the fit show strong evidence in favor of a two-Gaussian fit for the broad \ha{}.
We note that in our double-Gaussian model, the narrower of the two components has a dispersion of only $\sim230$\,\kms. Despite the narrow width, this component is at least twice broader than Gaussian components of forbidden lines, as listed in Table~\ref{table:linewidth}. Since it has no corresponding component in forbidden lines, we argue that this component is not a part of narrow emission lines but indeed originated from the broad line region.

Combining our updated broad \ha{} line width with the photometric time lag of $83\pm14$\,min based on the 2018 campaign by \citet{Woo+19}, we report the mass of the black hole in NGC 4395 as $(1.7\pm 0.3)\times10^4$\,\msun{} \citep[with the virial coefficient $f=4.47$ from][]{Woo+15} -- a factor of $\sim$2 larger than reported by \citet{Woo+19}. If we take the uncertainty in the virial coefficient into account \citep[i.e., $\sigma(\log_{10}
f) = 0.12$;][]{Woo+15}, the mass measurement should be read as $\verrud{1.7}{0.7}{0.5}\times10^4$\,\msun{} with the photometric lag. When we consider the constraint of $< 3$\,hr on the time lag from this spectroscopic campaign, we obtain the mass constraint of $< 4\times10^4$\,\msun{}.
While these are close to what \citet{Filippenko&Ho03} estimated via the BLR size-luminosity relation, it is significantly smaller than the \civ{} broad-line reverberation mapping result by \citet{Peterson+05}.
As \citet{Woo+19} discussed, the difference seems to originate not from the time-lag measurement, but from differences in the broad-line widths; \citet{Peterson+05} reported a \civ{} BLR size of 48--66\,min, while the line dispersion is estimated to be $\sigma_\mathrm{B \mbox{\ion{C}{4}}} \approx 2900$\,\kms\ based on the noisy root-mean-squared spectra. This emphasizes that the line width measurement plays a critical role in the mass determination of the AGN. We argue that the consistency of the broad \ha{} width between our work and that inferred from the data by
\citet{Woo+19} indicates that our \ha{} line-width measurement is reliable. In contrast, our time-lag measurement, as well as the aforementioned photometric time lags, is not significantly different from that of \citet{Peterson+05}.

One interpretation of the discrepancy of the mass is that the BLR structures for these two emission lines in NGC 4395 exhibit very large differences, resulting in different virial coefficients. If this is the case, the virial coefficient for \civ{} has to be larger than that of \hb\ by more than a factor of 20. However, based on a large sample of AGNs, a number of studies showed that the virial coefficients of the two emission lines are not dramatically different although the \civ{} is somewhat broader than \hb \citep[e.g.,][]{Vestergaard&Peterson06, Park+13, Park(DS)+17, Shen&Liu12, Williams+21} While recent work comparing the virial coefficient for \civ{} and Balmer lines \citep{Williams+20} does not find evidence for major differences in the virial coefficients, the situation could be different for NGC 4395 given the widely different masses. On the other hand, the \civ{} line from \citet{Peterson+05} exhibited an irregular profile which, combined with the noisy root-mean-squared spectra, can make the line-width measurement difficult.

Finally, while we did not conclusively measure a time lag, our results show that the \ha{} spectroscopic time lag during our 2019 campaign is broadly consistent with the photometric time lags we previously reported \citep{Woo+19, Cho+20}. It is possible that the time lag has changed and became much shorter than 30\,min, while judging from the shapes of CCFs it is unlikely to be longer than 3\,hr. Thus, we argue that the uncertainty in the mass measurement induced by the time lag is at most a factor of 3, although additional monitoring campaigns are needed to draw a firm conclusion. Based on these results, we find no significant evidence that NGC 4395 exhibits a departure from the $\mbh-\sigma_\ast$ relation, especially when considering the uncertainty of $\sigma_\ast$.

\subsection{Narrow-Line Flux as a Light-Curve Calibrator}

Narrow lines are often assumed to be invariant over the period of a reverberation-mapping campaign, so they can be used as a built-in calibrator for the flux and spectral shape of broad lines between different exposures, as demonstrated by  \citet{vanGroningen&Wanders92}. However, for a nearby AGN where the host galaxy and its NLR are typically resolved using ground-based telescopes, the variation of the flux contamination from the extended NLR can be significant depending on the seeing conditions and their variability throughout the campaign, thus invalidating the assumptions behind the method of \citet{vanGroningen&Wanders92}. As we demonstrated in \S~\ref{sss:nl_excess}, modeling the physical properties of narrow emission lines provides a way to correct for differential slit losses.

This source of systematic uncertainty can be eliminated by monitoring the AGN using an IFU spectrograph. The spatial information afforded by an IFS enables the integration of the spectra over a selective region of the sky, mitigating the seeing variation via PSF matching, and having more accurate flux calibration via simultaneous comparison-star observations if the FoV is sufficiently large to include one.

\section{Conclusions}

We performed spectroscopic and photometric monitoring of the intermediate-mass AGN in NGC 4395 over 3 nights in 2019, using the Gemini and Keck telescopes along with a number of 1\,m-class telescopes. The main results are summarized as follows.

(1) We detected significant variation of the narrow emission-line flux (i.e., \SII) among exposures in the slit-based spectra. We demonstrated that the variability is due to seeing-dependent contamination from the extended NLR, by investigating the relation between the \SII{} $\lambda 6717/\lambda 6731$ ratio and the total flux of the \SII{} doublet. By developing a method to correct for this effect, we presented the calibrated light curves of the broad \ha{}.

(2) By modeling the narrow lines (\NII, \ha, \SII) with two Gaussian components and the broad \ha\ with two additional Gaussian components, we improved the decomposition of emission lines and determined the line dispersion of broad \ha{} to be $\sigma_\mathrm{BH\alpha} = 586\pm19$\,\kms{}, which is 40\% larger than that reported by \citet{Woo+19}.

(3) While we obtained no conclusive spectroscopic lag between the AGN continuum and broad \ha{}, owing to a combination of insufficient variability structure and partly bad weather, we constrained the lag to be between 0 and 3\,hr, which is consistent with the photometric lag of $\sim 80$\,min determined by Woo et al. (2019).

(4) We reported the updated mass of the AGN in NGC 4359 as $(1.7^{+0.3}_{-0.3})\times10^4$\,\msun{}, by combining the revised line dispersion of \Ha{} and the photometric lag of \Ha{} \citep{Woo+19}. Even if we consider the upper limit of the time lag from our campaign, the mass should be smaller than $4\times10^4$\,\msun{}. Our estimated black hole mass is significantly different from that reported by \citep{Peterson+05} based on \civ. Reconciling the black hole mass estimates would require substantially different virial coefficients for the two emission lines. On the other hand, the difference between the virial coefficients of \civ{} and Balmer lines seems to be much smaller than the required amount. This tension highlights the importance of precise line-width measurements with sufficient signal-to-noise ratio for estimating the mass of black holes in AGNs.

\acknowledgments

%

% NRF (and HC personal acknowledgment)
This work has been supported by the Basic Science Research Program through the National Research Foundation of Korean Government (2021R1A2C3008486). The work of H.C. was supported by an NRF grant funded by the Korean Government (NRF-2018H1A2A1061365-Global Ph.D. Fellowship Program). H.C. would like to thank Yoo Jung Kim and Yoonsoo P. Bach for their help in the statistical discussion.
% NSF and other fundings.
T.T. acknowledges support from the Packard Foundation through a Packard Research Fellowship and (together with P.W.) by the U.S. National Science Foundation (NSF) through grant AST-1907208 (``Collaborative Research: Establishing the Foundations of Black Hole Mass Measurements of AGN Across Cosmic Time'').
Research at U.C. Irvine was supported by NSF grant AST-1907290. V.N.B. gratefully acknowledges assistance from NSF Research at Undergraduate Institutions (RUI) grant AST-1909297. Note that findings and conclusions do not necessarily represent views of the NSF. V.U acknowledges funding support from National Aeronautics and Space Administration (NASA) Astrophysics Data Analysis Program (ADAP) grant 80NSSC20K0450.
A.V.F. was financially supported by the TABASGO Foundation, the Christopher R. Redlich Fund, and the U.C. Berkeley Miller Institute for Basic Research in Science (where A.V.F. is a Senior Miller Fellow).
% Telescope acknowledgments
We thank the staffs of the observatories where data were collected for their assistance.
% Keck
Some of the data presented herein were obtained at the W. M. Keck Observatory, which is operated as a scientific partnership among the California Institute of Technology, the University of California, and NASA. The Observatory was made possible by the generous financial support of the W. M. Keck Foundation.
% Gemini
Based on observations obtained at the international Gemini Observatory, a program of NSF’s NOIRLab, which is managed by the Association of Universities for Research in Astronomy (AURA) under a cooperative agreement with the NSF on behalf of the Gemini Observatory partnership: the NSF (United States), National Research Council (Canada), Agencia Nacional de Investigaci\'{o}n y Desarrollo (Chile), Ministerio de Ciencia, Tecnolog\'{i}a e Innovaci\'{o}n (Argentina), Minist\'{e}rio da Ci\^{e}ncia, Tecnologia, Inova\c{c}\~{o}es e Comunica\c{c}\~{o}es (Brazil), and Korea Astronomy and Space Science Institute (Republic of Korea).
% Mauna Kea
The authors wish to recognize and acknowledge the very significant cultural role and reverence that the summit of Maunakea has always had within the indigenous Hawaiian community.  We are most fortunate to have the opportunity to conduct observations from this mountain.
% LCOGT
This work makes use of observations from the Las Cumbres Observatory global telescope network.
% IRAF (https://noirlab.edu/public/research/scientific-acknowledgments/#iraf)
IRAF was distributed by the National Optical Astronomy Observatory, which was managed by the Association of Universities for Research in Astronomy (AURA) under a cooperative agreement with the National Science Foundation.

\bibliography{bib.bib}

\end{document}